\documentclass[aps,prl,twocolumn,superscriptaddress,amsmath,amssymb,endfloats]{revtex4-2}
\usepackage{graphicx}
\usepackage{dcolumn}
\usepackage{bm}
\usepackage{natbib}
\usepackage{color}
\usepackage{amsmath}
\usepackage[]{hyperref}
\hypersetup{
    colorlinks=true,
    allcolors = {blue},
    }
\usepackage[normalem]{ulem}
\newcommand{\remove}[1]{}

\newcommand{\add}[1]{\textcolor{black}{#1}}
\begin{document}

\title{Conductance measurements cannot distinguish crossed Andreev reflection from elastic co-tunneling in normal--superconductor--normal junctions}
\author{E.S.~Tikhonov}\affiliation{Osipyan Institute of Solid State Physics, Russian Academy of
Sciences, 142432 Chernogolovka, Russian Federation}
\affiliation{\add{Condensed-Matter Physics Laboratory, HSE University, Moscow 101000, Russian Federation}}
\author{V.S.~Khrapai}
\affiliation{Osipyan Institute of Solid State Physics, Russian Academy of
Sciences, 142432 Chernogolovka, Russian Federation}
\affiliation{National Research University Higher School of Economics, 20 Myasnitskaya Street, 101000 Moscow, Russian Federation}
\email{dick@issp.ac.ru}

\maketitle

Feng et al. \cite{Feng2025Long} present transport measurements of  three-terminal normal-superconductor-normal (NSN) devices based on topological insulator nanowires. They measure the full differential conductance matrix, defined as $G_{ij} = \partial I_i/\partial V_j$, beyond the linear response regime as a function of various control parameters. Here $i,j \in [1,2]$ denote normal terminals biased by the voltages $V_1$ and $V_2$ with respect to the grounded superconducting terminal and $I_1$ and $I_2$ are the measured currents. The core idea is based on the assumption that a relative contribution of the two competing processes of crossed Andreev reflection (CAR) and elastic co-tunneling (ECT) to non-local conductance can be systematically tuned by the choice of a bias voltage combination. 

In Fig.~3 of their paper, Feng et al. observe indications that the CAR is enhanced for $V_1=V_2$, whereas the ECT is enhanced for $V_1=-V_2$, in analogy with the quantum dot based experiment~\citep{Bordin2023Tunable}, and use the former combination to investigate the gate voltage dependence of $G_{ij}$ in the remainder of the paper (Figs.~ 4, 5 and 6). Based on the sign of the non-local conductance Feng et al. make conclusions about the dominance of the CAR or ECT and the case of the dominant CAR is interpreted as the evidence of the unusually long-range crossed Andreev reflection. 

Here, we argue that the interpretation of the experiment is misleading in two respects. First, the bias voltages impact the non-local differential conductance randomly, rather than systematically, and the bias symmetry of the non-local conductance in Fig.~3 can be explained by a fine tuned self-gating effect. Second, the full knowledge of the $G_{ij}$ is insufficient to make conclusions about the relative values of the CAR and ECT probabilities, in particular on the dominance of one of them.

Our analysis is based on the scattering matrix formalism of Ref.~\cite{Anantram1996Current}, which may have shortcomings beyond the linear response regime but has a benefit of capturing the main physics of the transport experiment in a transparent way. For our purposes it is enough to consider the case of a single-mode device. We have:
\begin{subequations}
\begin{align}
I_i   &= G_{i1}V_1 + G_{i2} V_2 \label{eq1.1} \\
G_{ij} & = G_0\left(\delta_{ij}-T_{ij}^{ee} + T_{ij}^{he}\right) \label{eq1.2} 
\end{align}
\end{subequations}
where $G_0$ is the spinful conductance quantum, $e$ ($h$) stand for the electron (hole), and $T_{ij}^{\alpha\beta}$ is the transmission probability of a type $\beta$ quasiparticle from the terminal $j$ into the terminal $i$ as a type $\alpha$ quasiparticle ($\alpha,\beta \in [e,h]$). For example, the non-local ECT and CAR probabilities from the terminal 2 to the terminal 1 are denoted as $T_{12}^{ee}$ and $T_{12}^{he}$, respectively, where we define the sign of the currents inflowing the device via all terminals as positive, which gives the opposite sign of the non-local conductance as compared to~\cite{Feng2025Long}. 

Feng et al. suggest that the ECT process is enhanced for $V_1 = -V_2$ and the CAR is enhanced for $V_1 = V_2$, in analogy with experiments on quantum dot devices~\cite{Bordin2023Tunable}. Note, however, that  the latter experiments discuss the total current, which has a completely different behaviour as compared to the differential conductance measured in~\citep{Feng2025Long}. From the Eqs.~(\ref{eq1.1}-\ref{eq1.2}) we find for the two quantities related to the terminal 1:
\begin{subequations}
\begin{eqnarray}
 I_1 =& 2\left(T_{11}^{he}+T_{12}^{he}\right)G_0V \,\,\,&\text{for } V_1 = V_2=V \label{eq2.1}\\
 I_1 =& 2\left(T_{11}^{he}+T_{12}^{ee}\right)G_0V \,\,\,&\text{for } V_1 =-V_2=V \label{eq2.2}\\
 G_{12} =& \left(-T_{12}^{ee}+T_{12}^{he}\right)G_0 \,\,\,&\text{for any } V_1,\, V_2\label{eq2.3} 
 \end{eqnarray} 
\end{subequations}
Eqs.~(\ref{eq2.1}-\ref{eq2.2}) show that if $T_{11}^{he}=0$, \remove{as}\add{an additional constraint on the scattering matrix relevant for}\remove{in} quantum dots in the regime of sequential tunneling~\cite{Bordin2023Tunable}, the CAR and ECT contributions in $I_1$ can indeed be separated by choosing the proper bias combination. The $G_{12}$, however, is insensitive to the bias combination and the ECT/CAR competition holds in Eq.~(\ref{eq2.3}). This is natural, since during the $G_{12}$ measurement $V_1$ is fixed and only $V_2$ is varied, hence the bias correlation is lost. This shows that the hypothesis about the origin of the bias voltages impact on the non-local conductances raised in Ref.~\cite{Feng2025Long} is not justified and an alternative explanation of the data of Fig.~3 is due. We provide that explanation below.

\begin{figure}[t!]
\begin{center}
\vspace{0mm}
 \includegraphics[width=1\linewidth]{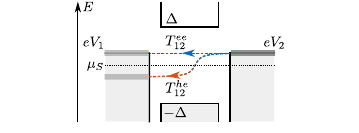}
  \end{center}
  \caption{The energy diagram of the scattering processes contributing to $G_{12}=\partial I_1/\partial V_2$ for  $V_1=V_2<0$ without the self-gating effect. A small modulation of $V_2$ is marked by the dark grey energy stripe. The modulation of the current $I_1$ occurs via quasiparticle flux in two light grey energy stripes, one via the ECT process ($T_{12}^{ee}$) and the other via the CAR process ($T_{12}^{he}$). $\mu_S$ and $\Delta$ denote, respectively, the chemical potential and the energy gap of the parent superconductor. In the presence of self-gating such an interpretation of the $G_{12}$ in terms of the CAR and ECT probabilities fails.} 
	\label{Fig1}
\end{figure}

Beyond the linear response regime the ECT and CAR probabilities modify in two ways~\cite{MaianiPRB2022}. One is due to the change of the quasiparticle energy $E$ and the other because of the bias-related change of the potential landscape (referred to as self-gating below). Thus the scattering probabilities acquire the energy and bias dependence of the form $T_{ij}^{\alpha\beta} = T_{ij}^{\alpha\beta}(E,V_1,V_2)$. The current $I_1$ is given by the integral over $E$ and the finite bias non-local conductance is obtained from its derivative $G_{12} = \partial I_1/\partial V_2$. Without the self-gating the bias dependencies drop from the $T_{ij}^{\alpha\beta}$ and one can still use the Eq.~(\ref{eq2.3}) with the CAR and ECT evaluated at $E=eV_2$, as shown in the following argument. Following~\citep{Anantram1996Current} for the total current in lead 1 we have:
\begin{equation}
I_1 = \frac{e}{h}\sum_{\alpha,\beta,j} \int \mathrm{sgn}_\alpha\left[\delta_{1j}\delta_{\alpha\beta}-T_{1j}^{\alpha\beta}(E)\right] f_{j,\beta}(E) dE, \label{App1}
\end{equation}
where $\mathrm{sgn}_e=1$, $\mathrm{sgn}_h=-1$ and the (electron and hole) quasiparticle energy distributions are expressed via the Fermi-Dirac function $f_0(E)$ as $f_{j,e} = f_0(E-eV_j)$ and $f_{j,h} = 1-f_0(-E-eV_j)$.
%
%
We find the differential conductance defined as $G_{12} =  \partial I_1/\partial V_2$ from a derivative of the Eq.~(\ref{App1}). Without self-gating the terms with $j=1$ drop. In the case of zero temperature we can use the following identities:
\begin{equation*}
\frac{\partial f_{2,e}}{\partial V_2} = e\delta(E-eV_2);\,\,\,\frac{\partial f_{2,h}}{\partial V_2} = -e\delta(E+eV_2).
\end{equation*}
where $\delta(x)$ is the Dirac delta-function. Finally, we use the electron-hole symmetry relations, which state $T_{12}^{ee}(E) = T_{12}^{hh}(-E)$ and $T_{12}^{he}(E) = T_{12}^{eh}(-E)$. Altogether we obtain for the differential conductance:
\begin{equation*}
G_{12} =  \frac{\partial I_1}{\partial V_2} = \frac{2e^2}{h}\left[T_{12}^{he}(E=eV_2)- T_{12}^{ee}(E=eV_2)\right],
\end{equation*}
which relates the $G_{12}$ with the energy dependent CAR and ECT probabilities, see Fig.~\ref{Fig1}. This shows that the dependence of the $G_{12}$ on $V_1$ cannot be explained without the self-gating. However, in the presence of self-gating the $G_{12}$ is no longer expressed via the CAR and ECT probabilities and additional terms are present, which further complicates the interpretation of the transport experiment.

The above analysis does not support the core idea of Feng et al. about the controllability of the ECT and CAR by the choice of the bias combination, which they believe is confirmed in Figs.~3d-g. The diagonal bias symmetry of $G_{12}$ on the $V_1$-$V_2$ plain observed in Fig.~3f and less ideally in Fig.~3e  is a result of self-gating and required a certain degree of fine tuning to suppress bias contributions of other symmetries. In Ref.~\citep{Feng2025Long} this fine tuning was achieved with the gate voltage ($V_\mathrm{g}$). Indeed, the data of Fig.~4 and Fig.~5 measured as a function of $V_\mathrm{g}$ for $V_1=V_2$ exhibit odd and even bias symmetry and varying sign of the non-local conductances randomly, mostly inconsistent with the core idea. This indicates that the ECT and CAR probabilities fluctuate strongly as a function of $E$ and $V_\mathrm{g}$.

Ref.~\cite{Feng2025Long} correlates the sign of the $G_{12}$ with a dominance of ECT or CAR, which is ambiguous in the presence of self gating. In particular, $G_{12}>0$, that is a negative sign non-local conductance as defined in Ref.~\citep{Feng2025Long}, may not imply that the CAR process is stronger than the ECT process. Even more importantly, the approach of Ref.~\cite{Feng2025Long} has a fundamental shortcoming that the relative strength of the CAR and ECT cannot be determined from the transport measurement alone. Indeed, it follows from the Eqs.~(\ref{eq1.1}-\ref{eq1.2}) that the matrix $G_{ij}$ remains the same if the local probabilities ($i=j$) are decreased by some amount $\delta T$ and the non-local probabilities ($i\neq j$) are simultaneously increased by the same amount. Obviously, beyond the linear response $\delta T$ can become an arbitrary function of $E$, $V_1$ and $V_2$. 

It is illuminating to see how the problem of relative strength of the CAR and ECT was solved in recent experiments in InAs nanowire based NSN devices~\cite{Denisov2021Charge, Denisov2022Heat}. In these experiments the $G_{ij}$ measurement was supplemented with the shot noise based measurement of the non-local thermal conductance $G_{th}\propto \sum\left( T_{12}^{he}+T_{12}^{ee}\right)$ (here $\sum$ denotes the sum over the eigenmodes). The $G_{th}$ showed regular $V_g$ behaviour with  negligible fluctuations, whereas the $G_{12}$ was much smaller and strongly fluctuated. This means that the CAR and ECT probabilities are nearly equal on average, $\sum T_{12}^{he}\approx \sum T_{12}^{ee}$, with a few percent accuracy~\citep{Denisov2021Charge}, and it is their minor variations that are responsible for the fluctuations of  $G_{12}$. We note that some  data in Ref.~\citep{Feng2025Long}, for example the green curves in Figs.~1e and 1g, closely resemble the behaviour of $G_{ij}$ observed earlier in Refs.~\cite{Denisov2021Charge, Denisov2022Heat}. This demonstrates that the conclusions of Ref.~\citep{Feng2025Long} about the dominance of the CAR or ECT drawn from the sign of the $G_{12}$ are highly speculative.

\section*{Acknowledgements}
We are grateful to Daniel Loss for a valuable discussion,  clarifications and criticism of our analysis. The work of E.S.T. was partially supported by the Basic Research Program of HSE. The authors declare no competing interests.

%

\end{document}